\documentclass[aps,twocolumn,showpacs,preprintnumbers,amsmath,amssymb,superscriptaddress,floatfix,nofootinbib]{revtex4}
\usepackage{graphicx}
\usepackage{epsfig}
\usepackage{epstopdf}
\usepackage{hyperref}
\usepackage{amsmath}
\allowdisplaybreaks[1]
\usepackage{amsfonts}
\usepackage{amssymb}
\usepackage{multirow}
\usepackage{makecell}
\usepackage[dvipsnames]{xcolor}
\usepackage{ulem}
\usepackage{longtable}
\usepackage{array}

\newcommand{\be}{\begin{equation}}
\newcommand{\ee}{\end{equation}}
\newcommand{\bea}{\begin{eqnarray}}
\newcommand{\eea}{\end{eqnarray}}
\newcommand{\bean}{\begin{eqnarray*}}
\newcommand{\eean}{\end{eqnarray*}}

\usepackage{subfigure}

\begin{document}

\title{The effective radius for production of baryon-antibaryon pairs from $\psi$ decays}

\author{Shu-Ming Wu}\email{wushuming@mail.itp.ac.cn}
\affiliation{CAS Key Laboratory of Theoretical Physics, Institute
	of Theoretical Physics,  Chinese Academy of Sciences, Beijing 100190, China}
\affiliation{University of Chinese Academy of Sciences (UCAS), Beijing 100049, China}

\author{Jia-Jun Wu}\email{wujiajun@ucas.ac.cn}
\affiliation{University of Chinese Academy of Sciences (UCAS), Beijing 100049, China}

\author{Bing-Song Zou}\email{zoubs@mail.itp.ac.cn}
\affiliation{CAS Key Laboratory of Theoretical Physics, Institute
	of Theoretical Physics,  Chinese Academy of Sciences, Beijing 100190, China}
\affiliation{University of Chinese Academy of Sciences (UCAS), Beijing 100049, China}
\affiliation{Synergetic Innovation Center for Quantum Effects and Applications (SICQEA),
	Hunan Normal University, Changsha 410081, China}

\begin{abstract}

By using the covariant L-S Scheme for the partial wave analysis, we deduce the ratios between the S-wave and D-wave contributions from the recent data of $\psi(1^-) \to B_8(1/2^+) \bar{B}_8(1/2^-)$ from the BESIII collaboration.
For the $J/\psi\to \Lambda\bar{\Lambda}$ and $J/\psi(\psi(2S))\to \Sigma^{+}\bar{\Sigma}^{-}$, the ratios are fixed and the average angular momenta are computed to estimate the effective radii of these processes.
The results show that the effective radii of these decays of $J/\psi(\psi(2S))$ are very small, which are around 0.04 fm.
Thus, it is a nice place to search excited baryon resonances with lower spin in the decays of $J/\psi(\psi(2S))$. 
Furthermore, for the other $\psi(1^-) \to B_8(1/2^+) \bar{B}_8(1/2^-)$ reactions, we propose some methods to get such effective radius.

\end{abstract}

\maketitle

\section{Introduction}


The study of the baryon resonance is one of the most important topic in the hadron physics.
The long standing puzzle of `` missing resonances '' is still unsolved~\cite{Klempt:2009pi,Crede:2013kia}.
A lot of excited nucleon resonances predicted in the quark model have not been observed in the experiments.
One possible reason is that these resonances may be too broad to observe, or the proper reactions which have significant signal of the resonances have not been explored experimentally yet.
Especially, the information of the nucleon resonances with mass heavier than 2 GeV is still very limited.


In the review of Particle Data Group~\cite{Zyla:2020zbs}, all of the nucleon resonances with mass larger than 2150 MeV are listed as follows, $N(2190)$, $N(2200)$, $N(2250)$,  $N(2300)$, $N(2570)$, $N(2600)$ and $N(2700)$. 
Among these nucleon resonances, the $N(2190)$, $N(2200)$, $N(2250)$, $N(2600)$, and $N(2700)$ resonances are found from experiments based on $\pi N,\gamma N$ reactions~\cite{Sokhoyan:2015fra, Svarc:2014zja,  Anisovich:2011fc, Arndt:2006bf, Cutkosky:1980rh, Hohler:1979yr}, while $N(2300)$ and $N(2570)$ are from the decay final states of $\psi(2S)$~\cite{Ablikim:2012zk}.
In the analysis of $\pi N$, $\gamma N$ reactions, Bonn-J\"ulich group~\cite{Ronchen:2012eg} and ANL-Osaka group~\cite{Kamano:2013iva} used dynamical coupled-channels approach to extract the $N^*$ resonances with various effective interactions.
Bonn-Gatchina group~\cite{Sarantsev:2009zz} and GWU/SAID group~\cite{Arndt:2006bf} used a K-matrix approach with Breit-Wigner resonances to analyze related channels. 
And the KSU group~\cite{Hunt:2018wqz} used a generalized energy-dependent Breit-Wigner parametrization of amplitudes to treat these channels.
The spins of all these $N^*$ resonances with mass heavier than 2.15~GeV are found to be larger than 5/2.
However, from the $\psi$ decay, the two new $N^*$ resonances found by BES collaboration from partial wave analysis (PWA) of $\psi' \to \bar{N} N^* + c.c. \to \bar{N}N\pi$ have spin less than or equal to 5/2.
It is interesting to find that the spins of the heavy $N^*$ resonances generated from $\pi N,\gamma N$ reaction are obviously larger than these from the $\psi$ decay.


In fact, the effective radii for these $N^*$ production processes play an important role for this kind of spin selection.
With larger effective radius ($r_{\text{eff}}$), the involved orbit angular momentum ($L$) should also be larger because of $L=r_{\text{eff}}\times p$, where $p$ is the relative momentum of two hadrons in the final states or the initial states.
Since the size of $N$ is around $1$ fm, the effective radius of the $\pi N$ and $\gamma N$ reaction will be around $1$ fm. 
For the $N^*$ with mass larger than 2 GeV, the momentum of $N$ is around $1$ GeV at the rest frame of the $N^*$.
Thus the angular momentum of $\pi N$ will be $L \sim r_{\text{eff}}\times p \sim 1 {\text{fm}}\cdot {\text{GeV}}\sim 5$.
For the $\pi N$ and $\gamma N$ reactions, the total angular momentum ($J$) of the initial state is from $L-1/2$ to $L+1/2$ and  $L-3/2$ to $L+3/2$, respectively.
It leads to the spins of the $N^*$ resonances generated from $\pi N$ and $\gamma N$ being roughly from $7/2$ to $13/2$.
Correspondingly, the spins of $N^*$ resonances observed from  $\pi N$ and $\gamma N$ reactions in Refs.~\cite{Sokhoyan:2015fra, Svarc:2014zja, Anisovich:2011fc, Arndt:2006bf, Cutkosky:1980rh, Hohler:1979yr} are all larger than $5/2$. 
However, in the $\psi$ decay, the $\bar{N}N^*$ only can be generated from the annihilation of the charm and anti-charm quarks, therefore, the effective radius is small, hence leads to a small orbit angular momentum between $\bar{N}N^*$.
Thus, qualitatively speaking, the spin of $N^*$ generated from the $\psi$ decay cannot be very large. Indeed only $N^*$ with low spins were observed by BES collaboration in their PWA of various $N^*$ production processes from $J/\psi$ and $\psi(2S)$ decays~\cite{Ablikim:2012zk,Bai:2001ua,Ablikim:2004ug,Ablikim:2009iw,Ablikim:2013vtm}.
%


To give a quantitative estimation of the upper limit for the spin of $N^*$ produced from $J/\psi$ and $\psi(2S)$ decays, it is important to study the effective interaction radius $(r_{{\text{eff}}})$ for these processes.
With $\bar{L}= r_{{\text{eff}}}\times p$, $r_{{\text{eff}}}$ can be determined by calculating the average orbital angular momentum $(\bar{L})$ and the relative momentum $(p)$. 
This article mainly studies the interaction radius of the $\psi(1^-) \to B_8 \bar{B}_8$.
%
%
In this work, we construct the amplitudes within the covariant L-S scheme ~\cite{Zou:2002yy, Dulat:2011rn, Anisovich:2004zz, Chung:1993da}, and then we do a PWA based on the BESIII data~\cite{Ablikim:2018zay, Ablikim:2020lpe, Ablikim:2017tys, Ablikim:2012eu, Ablikim:2018ttl,  Ablikim:2016sjb, Ablikim:2016iym}.
From the PWA, the ratio between the S-wave and D-wave contributions can be computed, and the  average orbital angular momentum $(\bar{L})$ of $B\bar{B}$ can be deduced.
At last, the effective ranges of these reactions can be obtained.


This article is organized as follows.
In Sec.~\ref{sec:formalism}, we present the theoretical framework of our calculation.
Our results are shown in  Sec.~\ref{sec:result} as well as some discussions.
Then a brief summary in Sec.~\ref{sec:summary} of this work is followed and Appendix \ref{AppendixA}, \ref{AppendixB} is presented at last.

\section{Theoretical framework} \label{sec:formalism}

The covariant L-S scheme is applied for studying the partial wave analysis of $\psi(1^-) \to B_8(1/2^+) \bar{B}_8(1/2^-)$ as follows~\cite{Zou:2002yy,Dulat:2011rn,Anisovich:2004zz,Chung:1993da}:
\begin{align}
	A 
	&=g_{S}\Psi^{(1)}_{\mu}\epsilon^{\mu}+g_{D}e^{i\delta}\Psi^{(1)}_{\mu}\epsilon_{\nu}\tilde{t}^{(2)\mu\nu},\label{eq:amp1}
\end{align}
where $g_{S}$ and $g_{D}$ are real number and have the same sign, stand for the S-wave and D-wave coupling constants, $\delta$ is the relative phase between S-wave and D-wave, $\epsilon_{\nu}$ is the polarization vector of $\psi$, $\Psi^{(1)}_{\mu}$ and $\tilde{t}^{(2)\mu\nu}$ are the covariant spin wave function $(\boldsymbol{S}_{B\bar{B}}=1)$ and D-wave orbital angular momentum covariant tensors, respectively.
Here the explicit expressions of them are as follows,
\begin{align}	
	\Psi^{(1)}_{\mu}&=\bar{u}_{B}(\gamma_{\mu}-\frac{r_{\mu}}{m_\psi + 2m_B})v_{\bar{B}},\\[2pt]
	\tilde{t}^{(2)\,\mu\nu}&=r^{\mu}r^{\nu} - \frac{r^2}{3}g^{\mu\nu} + \frac{r^2}{3m_\psi^2}p_{\psi}^{\mu}p_{\psi}^{\nu},
\end{align}
where $r=p_B-p_{\bar{B}}$,  $m_i$ and $p_i$ are the mass and four momentum of particle $i$, and $u_{B}(v_{\bar{B}})$ is the spinor wave function of baryon $B$($\bar{B}$). 
The detailed process is in Appendix~\ref{AppendixA}.

In the rest frame of the particle $\psi$, the differential decay rate of $\psi\to B\bar{B}$ is determined by the amplitude $A$ as:
\begin{align}
\frac{d\Gamma}{d\Omega}=\frac{1}{32\pi^2}\overline{|A|^2}\frac{|\vec{p}_{B}|}{m_{\psi}^2}.\label{eq:dgamma}
\end{align}

Because of the positron-electron collision, the $\psi$ spin projection is limited to be $\pm 1$ along the beam direction. 
With Eq.~(\ref{eq:dgamma}), the angular distribution for $\psi(1^-)\to B_8\bar{B}_8$ can be written as a function of the angle $\theta$ between $B$ direction and the electron beam as follows:
\begin{gather}
\frac{d\Gamma}{d \cos{\theta}}=C(1+\alpha \cos^2{\theta}),\\[4pt]
\alpha = \frac{-6(g_{D}/g_{S})^2|\vec{p}_B|^4 \!+\! 36 (g_{D}/g_S)|\vec{p}_B|^2\cos{\delta}}{9 \!+\! 10 (g_{D}/g_S)^2 |\vec{p}_B|^4 \!-\! 12(g_{D}/g_{S})|\vec{p}_B|^2\cos{\delta}},\label{eq:alpha1}
\end{gather}
where C is an overall normalization.

Furthermore, to determine the amplitude $A$ in Eq.(\ref{eq:amp1}), at least three inputs are needed to control the parameters, $g_S$, $g_D$ and $\delta$.
In the experimental side, not only $\Gamma$ and $\alpha$ are measured, but also another parameter $\Delta\Phi$ can be got by the cascade decay of $B$. 
The $\Delta\Phi$ is defined as the relative phase between the electric ($G_{E}^{\psi}$) and magnetic ($G_{M}^{\psi}$) form factors as follows,
\begin{align}
\Delta\Phi\equiv {\text{Arg}}(\frac{G_{E}^{\psi}}{G_{M}^{\psi}}).
\end{align}

By using $G_{E}^{\psi}$ and $G_{M}^{\psi}$, the amplitude of the decay $\psi(1^-) \to B_8(1/2^+) \bar{B}_8(1/2^-)$ can be re-defined as ~\cite{Faldt:2016qee,Faldt:2017kgy}:
\begin{gather}
\tilde{A} = -ie_{g}\bar{u}_BO_{\mu}v_{\bar{B}}\epsilon^{\mu}\label{eq:amp2},\\[4pt]
O_{\mu} = G_{M}^{\psi}\gamma_{\mu} - \frac{2m_B}{r^2}(G^{\psi}_{M}-G^{\psi}_{E})r_{\mu}\label{eq:gegmO},
\end{gather}
where $e_g$ is an overall couping constant.

By comparing two amplitudes defined in Eqs.(\ref{eq:amp1}, \ref{eq:amp2}), the $\Delta\Phi$ can be expressed as:
\begin{gather}\label{eq:tran}
	e^{i\Delta\Phi}= \sqrt{\frac{1+\alpha}{1-\alpha}}\left(\frac{2m_B}{m_\psi}-\frac{r^2}{m_\psi(m_\psi+2m_B)}\right)\frac{3+2r^2\boldsymbol{y}}{3-r^2\boldsymbol{y}},
\end{gather}
where $\boldsymbol{y}\equiv e^{i\delta}g_{D}/g_{S}$, $r^2=-4|\vec{p}_{B}|^2$.
We have derived them in Appendix~\ref{AppendixB} in detail.
Now through Eqs.(\ref{eq:alpha1}, \ref{eq:tran}), the $g_D/g_S$ and $\delta$ can be calculated from the measured $\alpha$ and $\Delta\Phi$.

The percentage of S-wave contribution can be estimated as the ratio between pure S-wave partial width and total width, $\Gamma_{S}/\Gamma_{\text{Total}}$, which is determined by the value of $g_D/g_S$:
\begin{align}
\frac{\Gamma_{S}}{\Gamma_{Total}}&=\frac{9}{9 + 2(g_D/g_S)^2 ( m_\psi^2 - 4 m_B^2 )^2}.
\end{align}

Then, the average angular momentum $\bar{l}$ can be approximated to
\begin{align}
\bar{l}&=l_S\frac{\Gamma_{S}}{\Gamma_{Total}} + l_D\frac{\Gamma_{D}}{\Gamma_{Total}}\nonumber\\
&=\frac{4(g_D/g_S)^2 ( m_\psi^2 - 4 m_B^2 )^2}{9 + 2(g_D/g_S)^2 ( m_\psi^2 - 4 m_B^2 )^2}.
\end{align}

At last, the effective decay range $r_{{\text{eff}}}$ can be estimated from $\bar{l}=r_{{\text{eff}}}\times p$.

\section{Results and discussions} \label{sec:result}

For the decays of $J/\psi\to\Lambda\bar{\Lambda}$ and $\Sigma^+\bar{\Sigma}^-$, $\psi(2S)\to\Sigma^+\bar{\Sigma}^-$, the parameters $\alpha$ and $\Delta\Phi$ have been both well measured~\cite{Ablikim:2018ttl,Ablikim:2020lpe} as shown in Table~\ref{table:result1}. 
The ratio between S-wave and D-wave coupling constants, $g_D/g_S$ (GeV$^{-2}$), and the relative phase, $\delta$, can be determined in our calculation as shown in Table~\ref{table:result1}. 
It shows that in the all three reactions the S-wave contribution is larger than $85\% $, and the effective radius of the $\psi$ decay is around $0.04$ fm.

\begin{table*}
	\caption{\label{table:result1}The parameters of $J/\psi\to\Lambda\bar{\Lambda}$,$\Sigma^+\bar{\Sigma}^-$, $\psi(2S)\to\Sigma^+\bar{\Sigma}^-$ reactions are including the measured $\alpha$ and $\Delta\Phi$(radian), the computed ratio between the coupling constants of S-wave and D-wave ($g_D/g_S$ (GeV$^{-2}$)), the relative phase between S-wave and D-wave ($\delta$), the ratio between partial width of S-wave and total width ($\Gamma_S/\Gamma_{\text{Total}}$), and the effective radius ($r_{\text{eff}}$(fm)).}
	\renewcommand\arraystretch{1.5}
	\begin{tabular}{|c|c|c|c|c|c|c|}
		\hline
		\multirow{3}*\text{Mode} & $\alpha$ & $\Delta\Phi$ & $g_D/g_S$ & $\delta$ & $\Gamma_{S}/\Gamma_{\text{Total}}$ & $r_{{\text{eff}}}$ \\
		\hline
		$J/\psi\to\Lambda\bar{\Lambda}$ & $0.461\pm0.013$\cite{Ablikim:2018zay} & $0.74\pm0.019$\cite{Ablikim:2018zay} & $0.180\pm0.005$ & $-0.804\pm0.024$ & $85.7\pm0.6\%$ & $0.0488\pm0.0021$ \\
		\hline
		$J/\psi\to\Sigma^+\bar{\Sigma}^-$ & $-0.508\pm0.010$\cite{Ablikim:2020lpe} & $-0.270\pm0.021$\cite{Ablikim:2020lpe} & $0.171\pm0.006$ & $2.67\pm0.04$ & $90.9\pm0.6\%$ & $0.0362\pm0.0024$ \\
		\hline
		$\psi(2S)\to\Sigma^+\bar{\Sigma}^-$ & $0.682\pm0.041$\cite{Ablikim:2020lpe} & $0.379\pm0.084$\cite{Ablikim:2020lpe} & $0.097\pm0.009$ & $-0.33\pm0.10$ & $88.3\pm2.0\%$ & $0.033\pm0.006$ \\
		\hline
	\end{tabular}
\end{table*}

\begin{figure*}[htbp]
	\centering
	\includegraphics[width=0.3\linewidth]{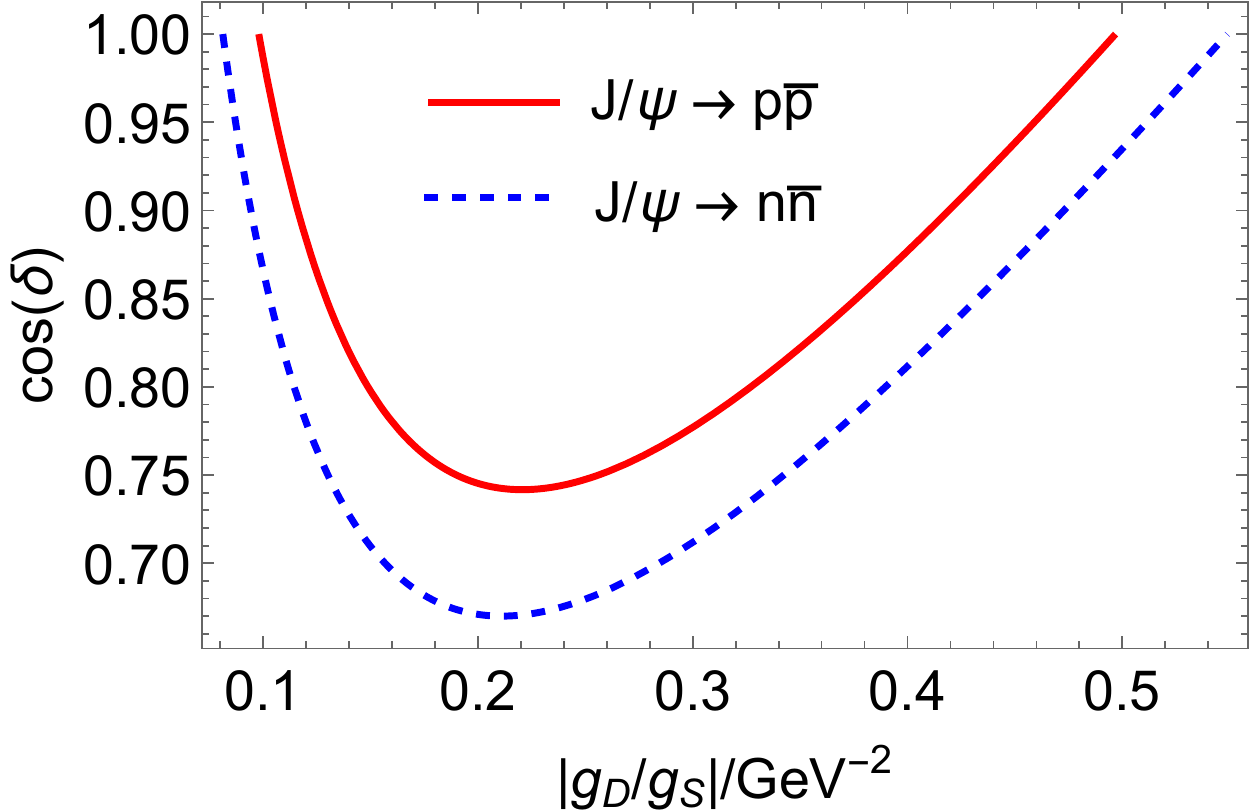} 
	\includegraphics[width=0.3\linewidth]{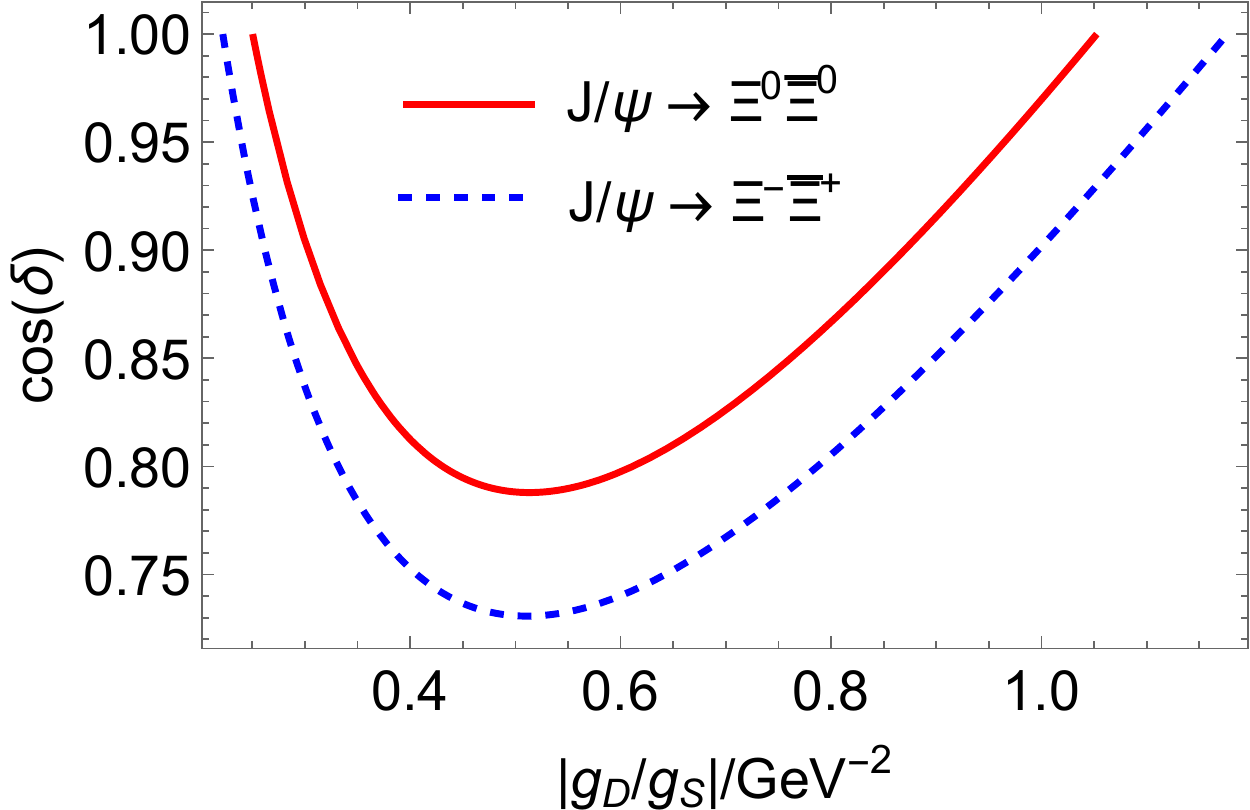}
	\includegraphics[width=0.3\linewidth]{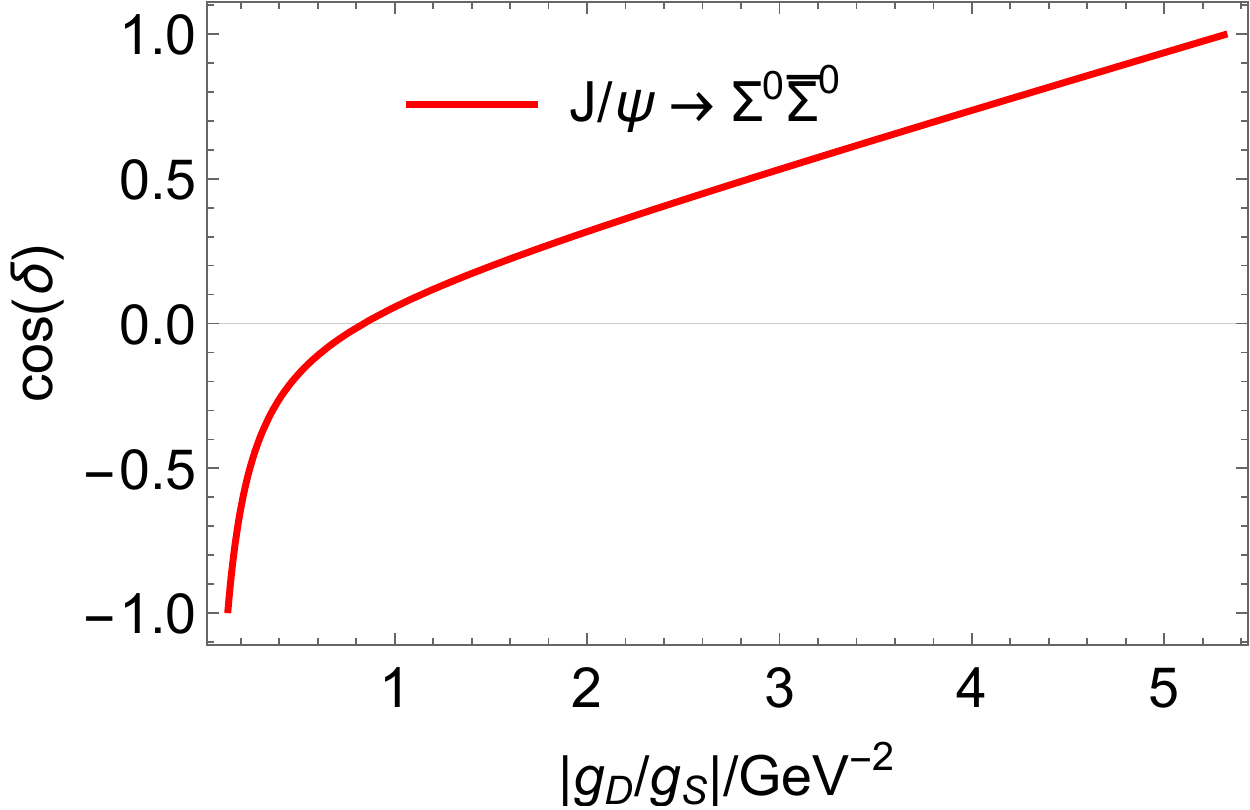}
	\\	
	\vspace{0.2in}
	\includegraphics[width=0.3\linewidth]{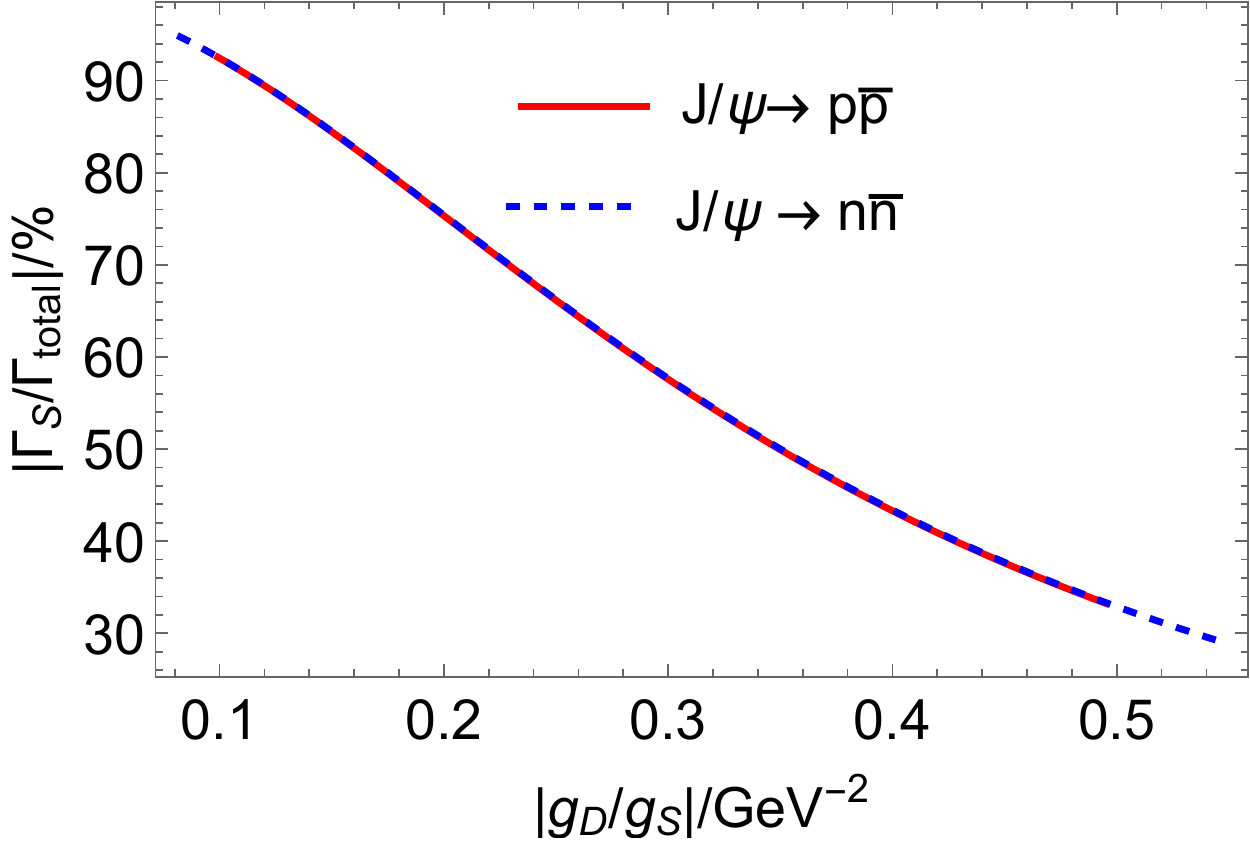} 
	\includegraphics[width=0.3\linewidth]{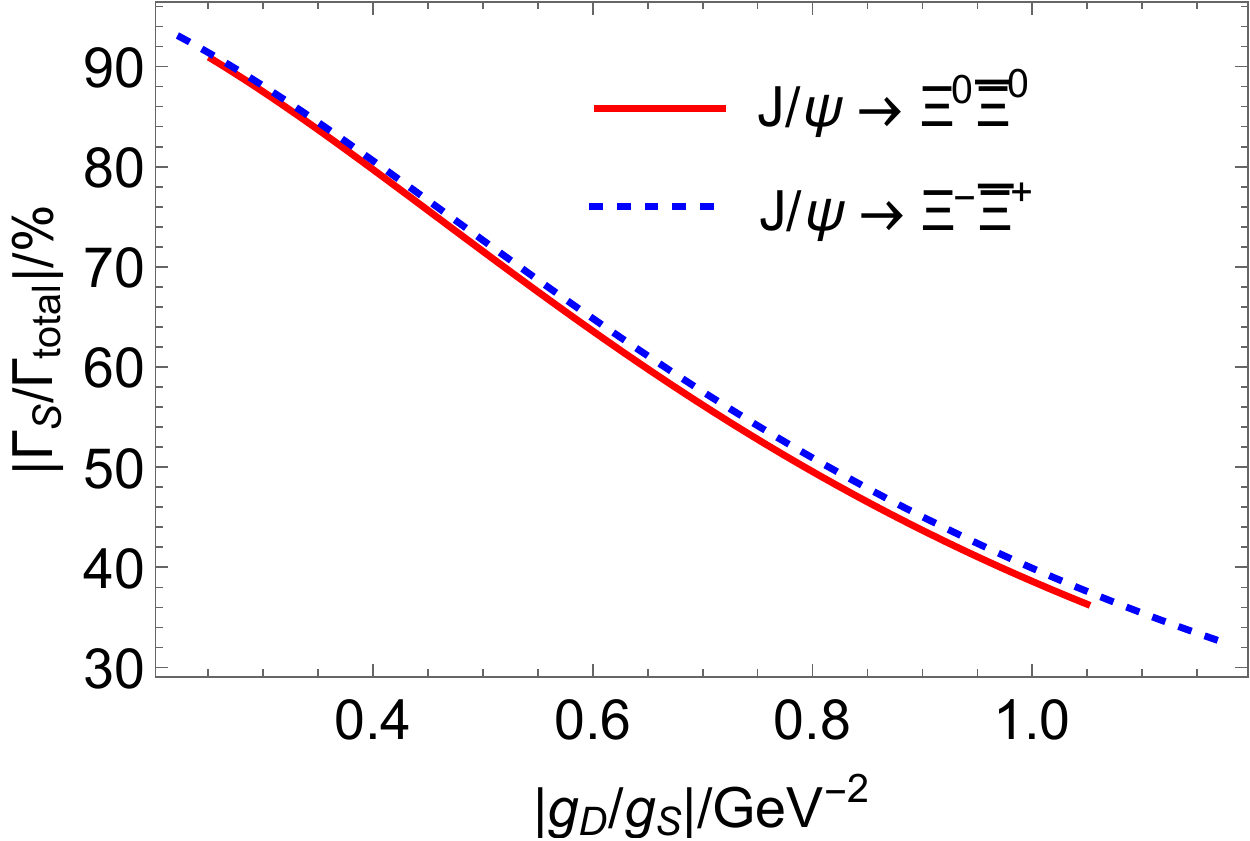}
	\includegraphics[width=0.3\linewidth]{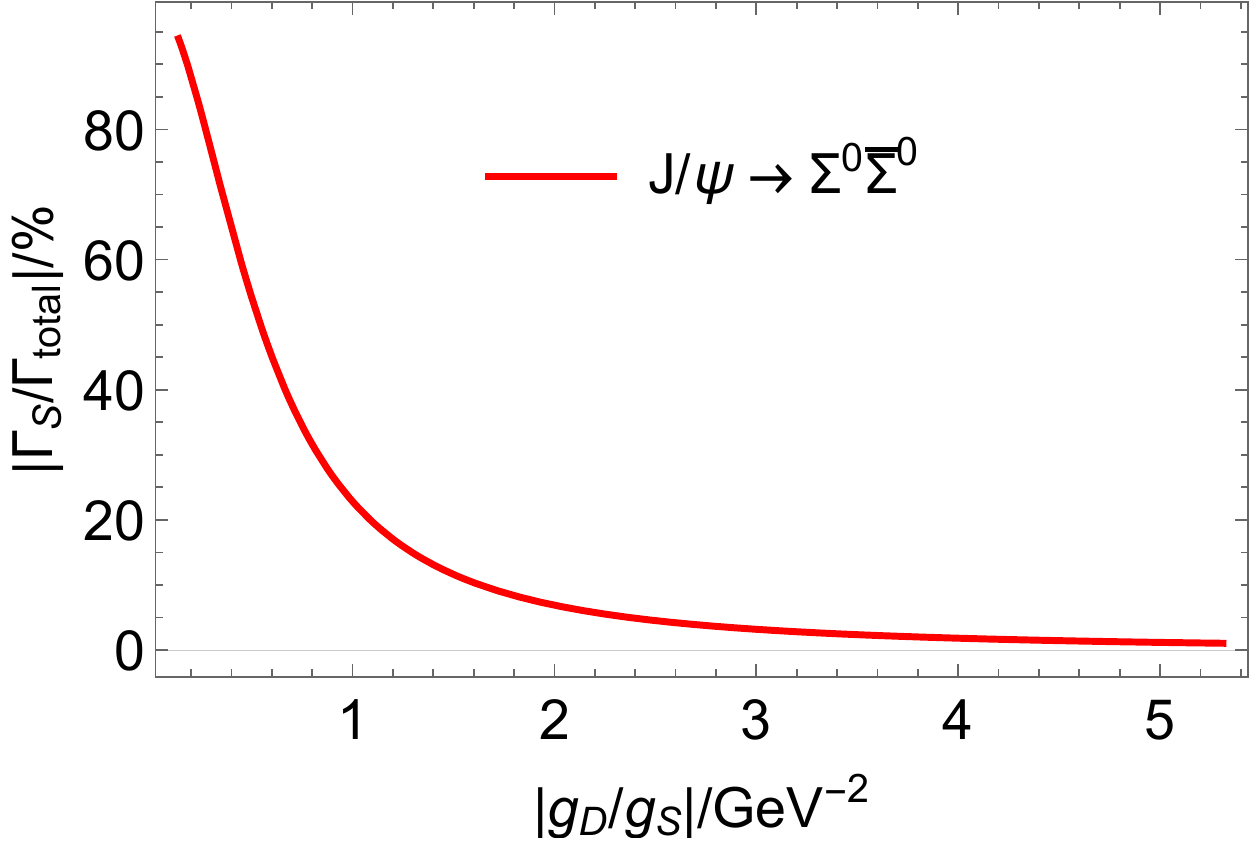} \\
	\caption{\label{fig:result3}The relationship between $g_D/g_S({\text{GeV}}^{-2})$ and $\delta$ in $J/\psi\to B_8\bar{B}_8$ with the fixed value of $\alpha$ from experiment.}
\end{figure*}

\begin{figure*}[htbp]
	\centering
	\includegraphics[width=0.3\linewidth]{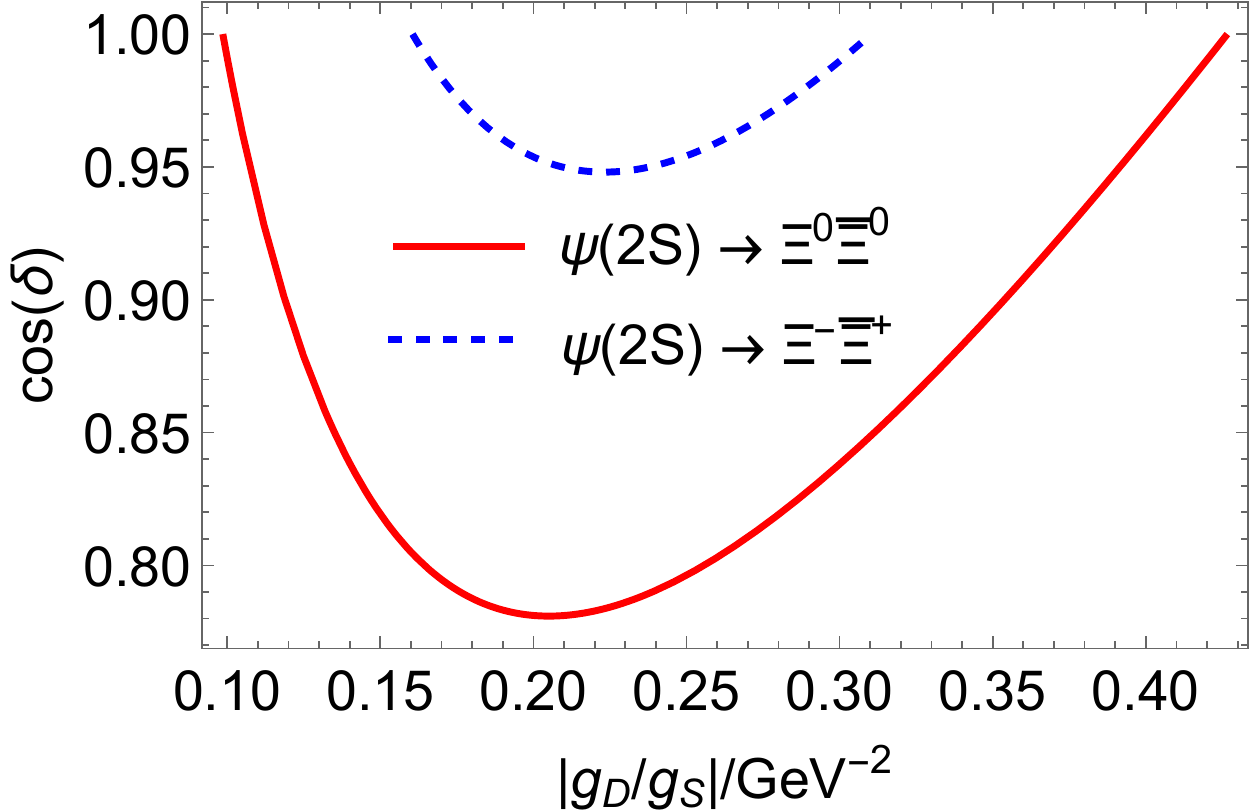} 
	\includegraphics[width=0.3\linewidth]{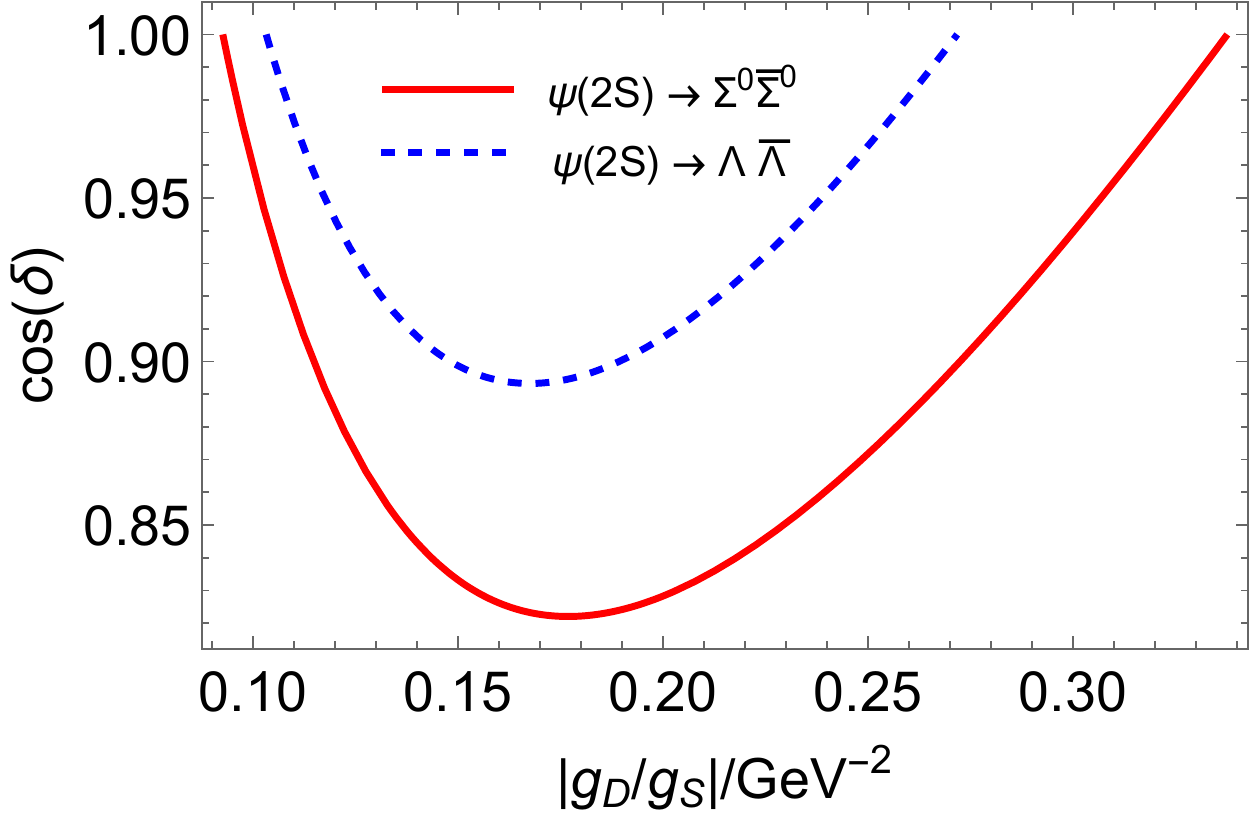}
	\includegraphics[width=0.3\linewidth]{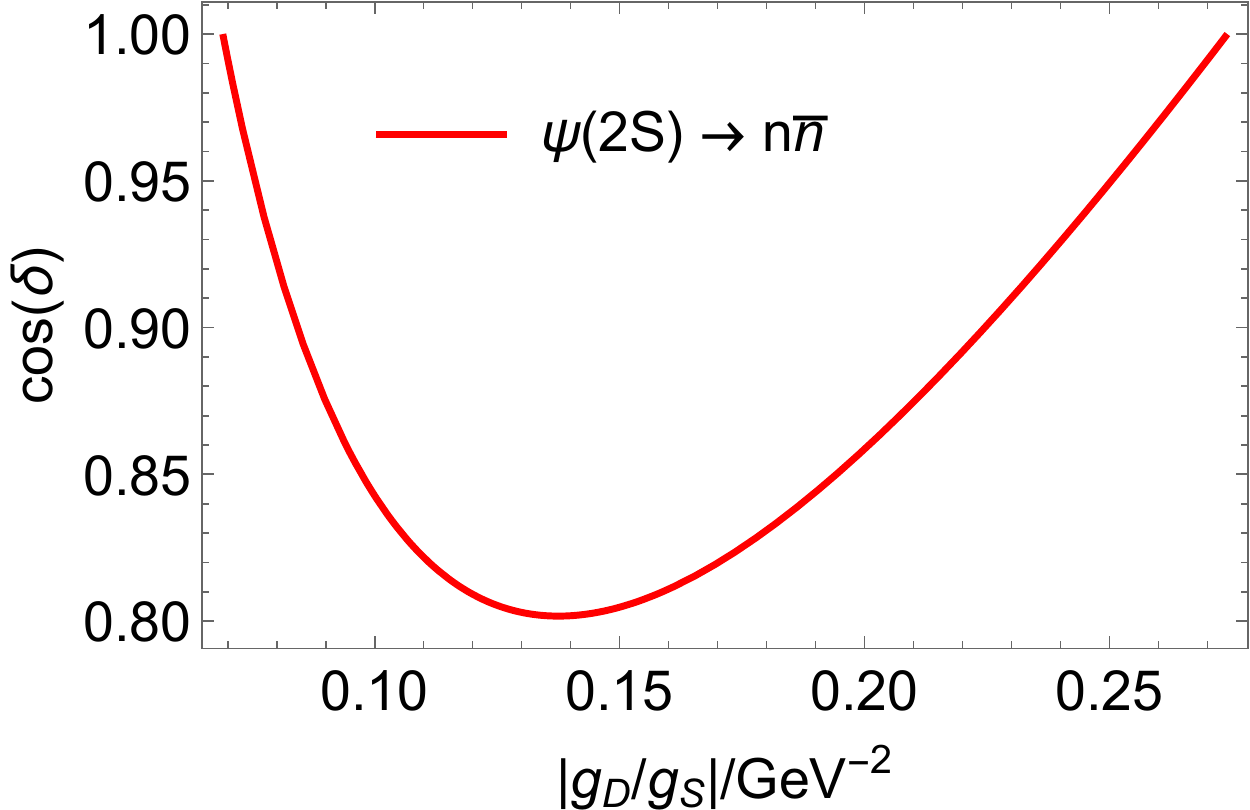}
	\\	
	\vspace{0.2in}
	\includegraphics[width=0.3\linewidth]{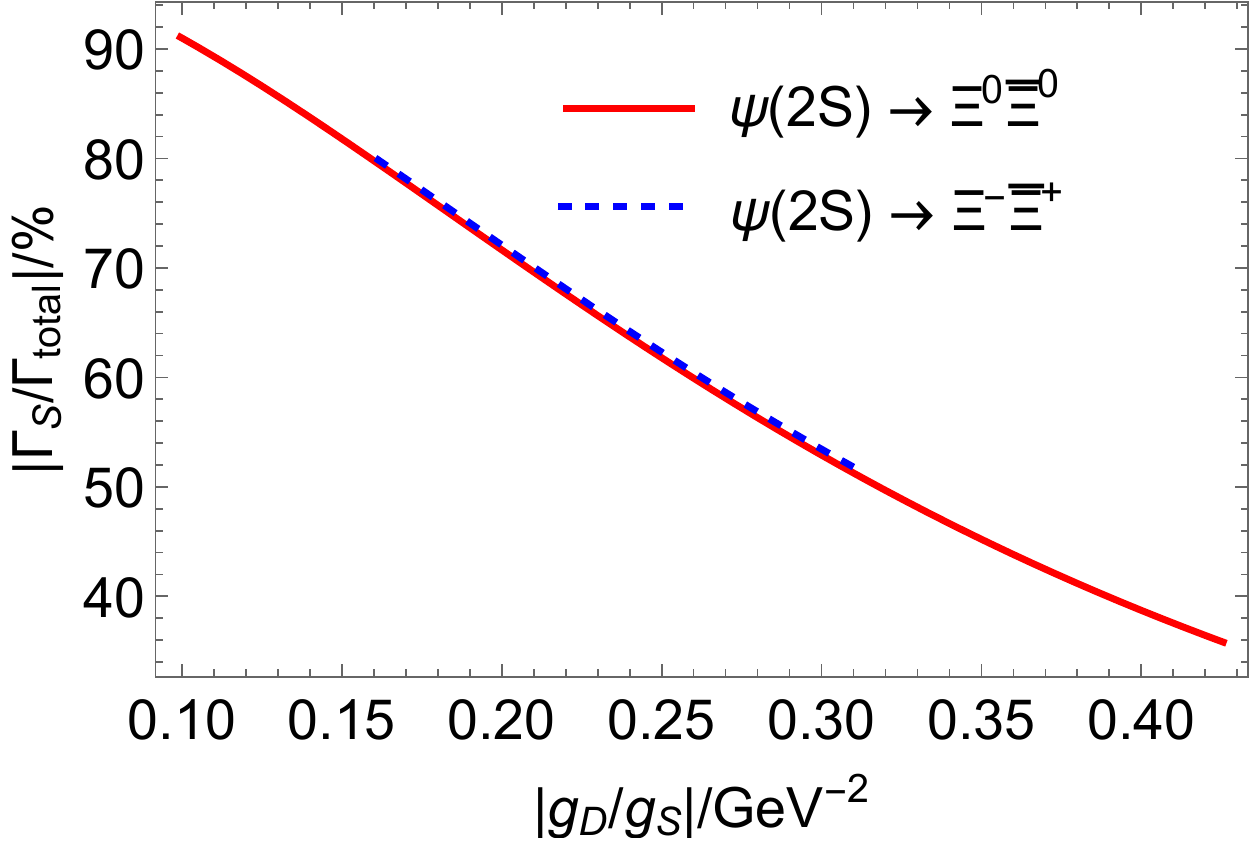} 
	\includegraphics[width=0.3\linewidth]{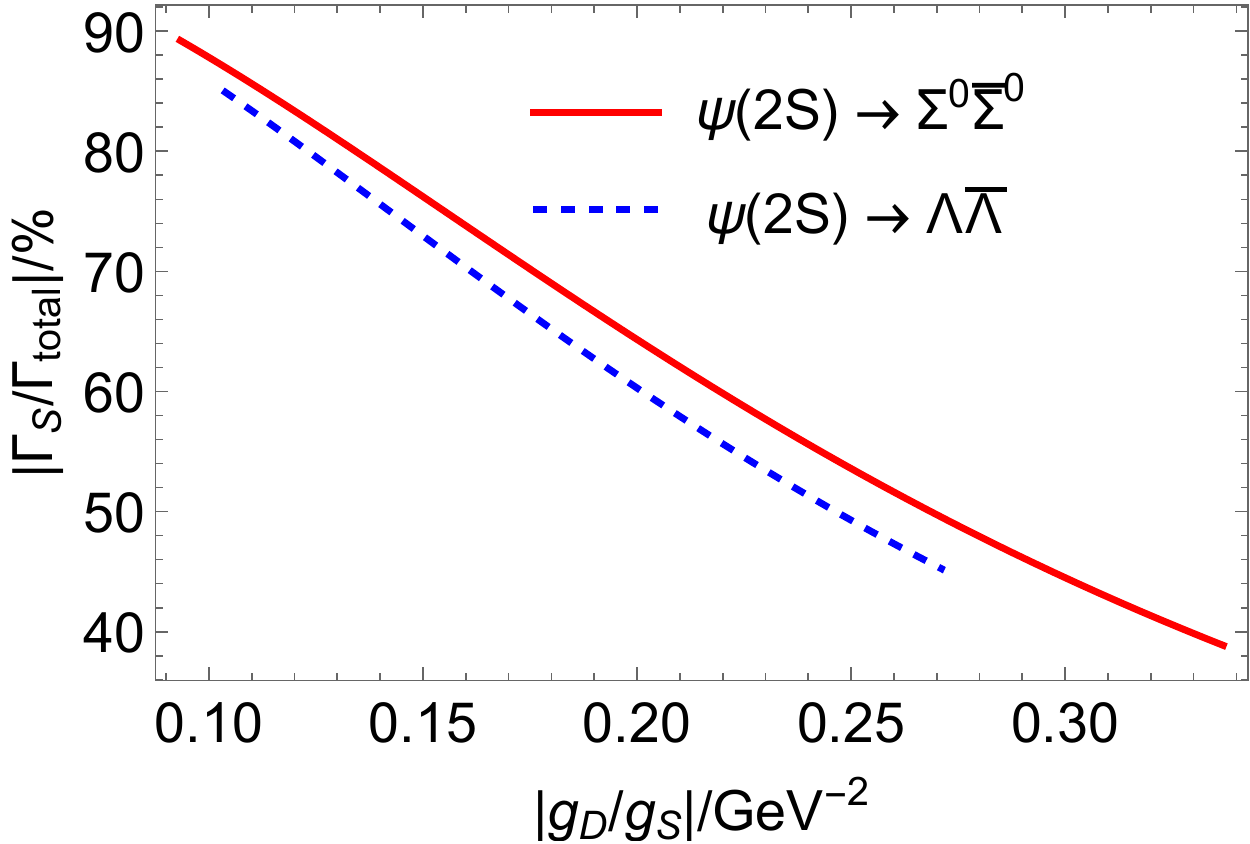}
	\includegraphics[width=0.3\linewidth]{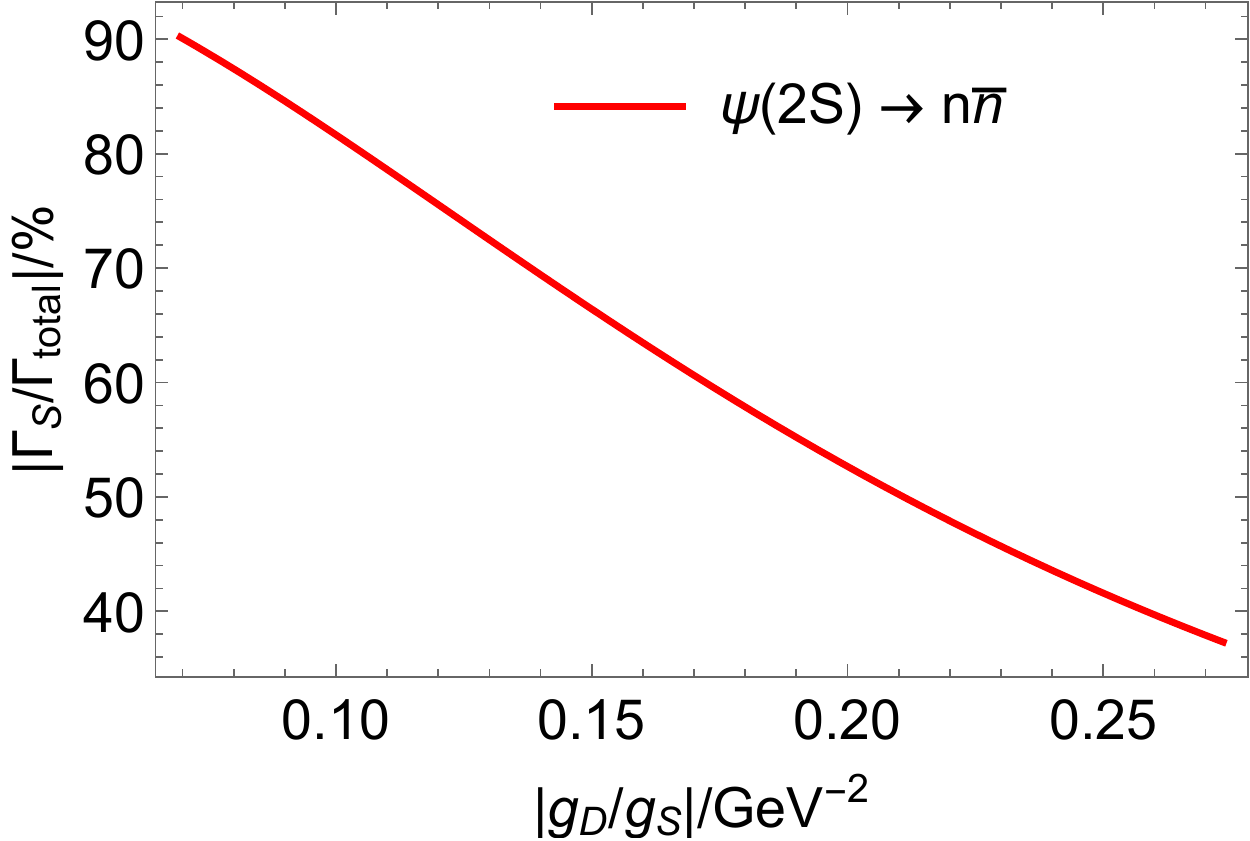} \\
	\caption{\label{fig:result4}The relationship between $g_D/g_S({\text{GeV}}^{-2})$ and $\delta$ in $J/\psi\to B_8\bar{B}_8$ with the fixed value of $\alpha$ from experiment.}
\end{figure*}

For the other decay channels, the $\Delta \phi$ have not been measured, but we still can estimate the range of the $r_{{\text{eff}}}$ from the measured $\alpha$. 
At first, by using Eq.(\ref{eq:alpha1}) and the value of $\alpha$, the relationship between $g_D/g_S$ (GeV$^{-2}$) and $\delta$ will be fixed as shown in the Fig.\ref{fig:result3} and Fig.\ref{fig:result4}. 
From Fig.\ref{fig:result3} and Fig.~\ref{fig:result4}, we find the possible values of $g_D/g_S$ (GeV$^{-2}$) are limited in a range.
It will lead to a range of the $r_{{\text{eff}}}$ as shown in Table \ref{table:result2}, and the values are all below the 0.4 fm which is obviously much smaller than the size of proton.
Especially, for the $\alpha$ of $\psi(2S)\to p\bar{p}$, there are $\alpha=1.03\pm0.09$, however the value of $\alpha$ should satisfy $|\alpha|\le 1$. 
Thus, it is failed to extract any information from the current value of the $\alpha$ of $\psi(2S)\to p\bar{p}$.

\begin{table*}
	\caption{\label{table:result2}The $\alpha$ and the estimation of $r_{{\text{eff}}}$(fm) in the reactions of $\psi \to B_8\bar{B}_8$ without $\Delta\Phi$ measurement.}
	\renewcommand\arraystretch{1.5}
	\begin{tabular}{|c|c|c|c|c|c|}
		\hline
		\multirow{3}*\text{Mode} & $\alpha$ & the range of $r_{{\text{eff}}}$(fm) & $\psi_{2S}\to\Lambda\bar{\Lambda}$ & $0.82\pm0.10$\cite{Ablikim:2017tys} & $[0.040-0.148]$ \\
		\hline
		$J/\psi\to p\bar{p}$ & $0.595\pm0.027$\cite{Ablikim:2012eu} & $[0.023-0.214]$ & $\psi_{2S}\to p\bar{p}$ & $1.03\pm0.09$\cite{Ablikim:2018ttl} & \\
		\hline
		$J/\psi\to n\bar{n}$ & $0.50\pm0.25$\cite{Ablikim:2012eu} & $[0.016-0.228]$ & $\psi_{2S}\to n\bar{n}$ & $0.68\pm0.23$\cite{Ablikim:2018ttl} & $[0.024-0.156]$\\
		\hline
		$J/\psi\to\Sigma^0\bar{\Sigma}^0$ & $-0.449\pm0.028$\cite{Ablikim:2017tys} & $[0.022-0.396]$  & $\psi_{2S}\to\Sigma^0\bar{\Sigma}^0$ & $0.71\pm0.15$\cite{Ablikim:2017tys} & $[0.030-0.172]$ \\
		\hline
		$J/\psi\to\Xi^0\bar{\Xi}^0$ & $0.66\pm0.08$\cite{Ablikim:2016sjb}  & $[0.044-0.308]$ & $\psi_{2S}\to\Xi^0\bar{\Xi}^0$ & $0.65\pm0.23$\cite{Ablikim:2016sjb} & $[0.027-0.196]$ \\
		\hline
		$J/\psi\to\Xi^-\bar{\Xi}^+$ & $0.58\pm0.12$\cite{Ablikim:2016iym}  & $[0.034-0.331]$ & $\psi_{2S}\to\Xi^-\bar{\Xi}^+$ & $0.91\pm0.27$\cite{Ablikim:2016iym} & $[0.061-0.148]$ \\
		\hline
	\end{tabular}
\end{table*}

From the above results, the effective radius of $\psi$ decay is very small. 
As discussed in the introduction, the small effective radius will lead to the small orbital angular momentum of the process of $\psi\to \bar{N}N^*$, resulting in the $N^*$ produced by this process to be dominated by the low spin $N^*$. 
Therefore, when dealing with $N^*$ partial wave analysis in $\psi$ decay, the highest partial wave can be taken as an angular momentum $l=5/2$.
Furthermore, some decay channels have not enough experimental inputs, only a range of $r_{\text{eff}}$ is estimated. 
It is still necessary to confirm the effective reaction radii of them, which will accurately tell us how large spin of the resonance can be found in the decay of $\psi$.

\begin{figure}[htbp]
	\centering
	\includegraphics[width=0.75\linewidth]{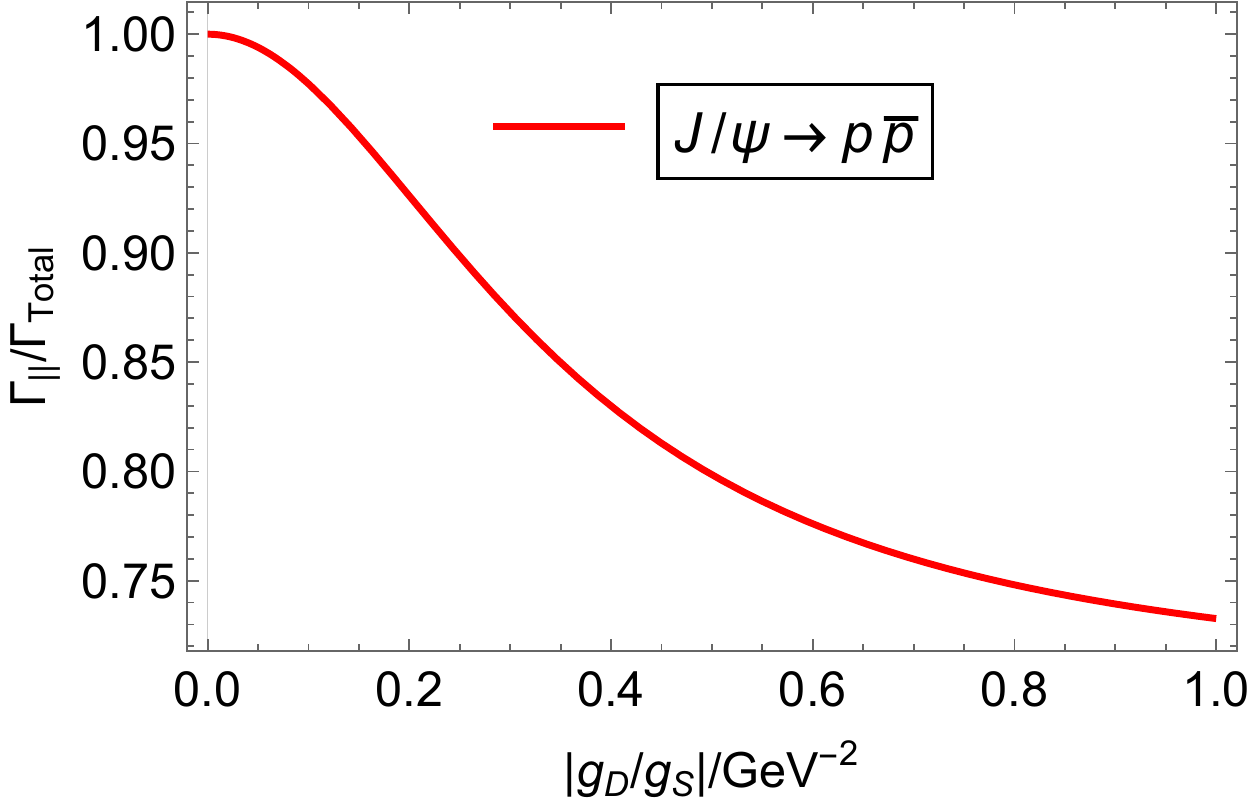} 
	\caption{\label{fig:result5}The ratio between $\Gamma_{\parallel}$ and $\Gamma_{{\text{Total}}}$ vs the $g_D/g_S({\text{GeV}}^{-2})$ in $J/\psi\to p\bar{p}$.}
\end{figure}

Obviously, more observables are needed to be measured if we want to fix the effective ranges of all decay channels. 
For the $\Sigma\bar{\Sigma}$, $\Xi\bar{\Xi}$ and $\Lambda\bar{\Lambda}$ decay channels, the  $\Delta\Phi$ can be got from the measurement of differential decay of cascade decay.
For example, from the data of the final states' angular distribution of decay of $J/\psi\to \Xi\bar{\Xi}\to \Lambda\pi\bar{\Lambda}\pi$, the $\Delta\Phi$ of this reaction can be measured. 
However, for the decay of $J/\psi$ or $\psi(2S)\to p\bar{p}$ or $n\bar{n}$, there are no cascade decay information of proton or neutron, since proton is stable and the life time of neutron is about 15 minutes, respectively. 
As a result, the $\Delta\Phi$ of these two reactions can not be obtained easily. 
It is worthy to mention that the polarization information of $p\bar{p}$ or $n\bar{n}$ can provide completely control of the $g_D/g_S$(Gev$^{-2}$). 
We can define $\Gamma_{\parallel}$ to stand for the decay width of the process where $p(n)$ and $\bar{p}(\bar{n})$ have the same polarization:
\begin{gather}
\Gamma_{\parallel}=\frac{|\vec{p}_{B}|}{32\pi^2m_{\psi}^2}\int\frac{1}{2}\sum_{s_\Psi,s_B=s_{\bar{B}}}|A(s_\Psi,s_B,s_{\bar{B}})|^2d\Omega\nonumber\\
=\Gamma_S+0.7\Gamma_D.
\end{gather}
In other word, once the $\Gamma_{{\text{Total}}}$ and $\Gamma_{\parallel}$ are both fixed, the $\Gamma_S$ and $\Gamma_D$ can be computed. 
In Fig.\ref{fig:result5}, it describes the relation between the $\Gamma_{\parallel}/\Gamma_{{\text{Total}}}$ and the $g_D/g_S$(GeV$^{-2})$. 

In all of the reactions mentioned above, the phase ($\delta$) between S-wave and D-wave plays an important role.
Generally speaking, if we assume that the $B_8\bar{B}_8$ is generated from the $c\bar{c}$ annihilation through three gluons, to the leading order, the relative phase $\delta$ should be zero. 
However, there are a lot of other mechanisms which will provide different phase between the S-wave and D-wave, a free parameter for relative phase is acceptable. 
In our Tables~\ref{table:result1} and \ref{table:result2}, all of the values of the $\alpha$ for different processes are listed and among these data we can find only the values of $\alpha$ for $J/\psi \to \Sigma^{+}\bar{\Sigma}^{-}$ and $\Sigma^0\bar{\Sigma}^0$ are negative while others are all positive.
It implies that there is a relative phases ($\delta$) difference of $\pi$ between the $J/\psi \to \Sigma^+\bar{\Sigma}^-(\Sigma^0\bar{\Sigma}^0)$ processes and others as shown in Table~\ref{table:result1}. 

From the other point of view, the electric and magnetic form factors expressing the amplitude as shown in Eqs.(\ref{eq:amp2}-\ref{eq:gegmO}) can also reflect the value of $\alpha$ and branching ratio ($Br$) as follows,
\begin{align}
Br&=|G_E|^2+|G_M|^2,\\
\alpha&=\frac{|G_M|^2-2|G_E|^2}{|G_M|^2+2|G_E|^2}.
\end{align} 
Then from the branching ratios and the values of $\alpha$ for each channel, their $|G_E|$ and $|G_M|$ can be computed as shown in Table~\ref{table:resultGEGM}.
As described in Ref.~\cite{Ferroli:2018yad}, the sign difference of $\alpha$ between $\Sigma^0\bar{\Sigma}^0$ and $\Lambda\bar{\Lambda}$ can be explained by using a SU(3) symmetry breaking model~\cite{Ferroli:2019nex}.
In Ref.~\cite{Ferroli:2019nex}, they provide all amplitudes listed in Table ~\ref{table:resultAmp}, where $G_0$ is related to the three gluons interaction within SU(3) symmetry, $D_e$ and $F_e$ for the EM breaking effects, $D_m$ and $F_m$ describe the mass difference breaking effects, $\phi$ is relative phase between strong interaction and EM interaction, $R$ is the ratio between $gg\gamma$ and $ggg$ contributions. 
By following the method of Ref.~\cite{Ferroli:2018yad}, we can decompose the electric and magnetic amplitudes as the different parts, respectively.
At last for electric and magnetic amplitudes, we can solve them independently, the results are shown in Table~\ref{table:resultpart}.
From these solutions, we can find that the $g_0$ of electric amplitudes is even smaller than that of other symmetry breaking terms, while $g_0$ is dominant in the magnetic amplitudes.
Of course such parameters are not proper, especially for the electric amplitudes. 
However, in Ref.~\cite{Ferroli:2019nex},  this model is able to explain the branching ratios since most channels are dominated by the magnetic amplitudes.
In Ref.~\cite{Ferroli:2018yad}, they use this model explain the $\alpha$ for the $\Sigma^0\bar{\Sigma}^0$ and $\Lambda\bar{\Lambda}$, but once all octets baryon channels are considered it is not so reasonable to explain the values of $\alpha$ which need electric and magnetic amplitudes both well described.
Thus, beside the SU(3) breaking effect mentioned above,  other mechanisms are needed to solve this problem. 
For example, for the $\Sigma$ the total spin and isospin of light quark pair are both 1,  {\sl i.e.},  so called “bad diquark”~\cite{Dobbs:2017hyd,Jaffe:2003sg}, while for the $\Lambda$ the [ud] quark pair has both spin and isospin to be 0, {\sl i.e.},  so called “good diquark”.
It may be a reason causing the angular distribution difference for the case of  $\Sigma\bar{\Sigma}$ from others. 
The hadronic loops may be another possible mechanism.
The study of these possible mechanisms are beyond the scope of this paper and should be explored in the future.
An important point is :  although the detailed production mechanisms may be different to cause the difference of angular distributions, the deduced small effective radii for their production from vector charmonium annihilation are nearly the same and lead to the S-wave dominant final states for both cases.

\begin{table}
	\caption{\label{table:resultGEGM}The values of $|G_E|^2$ and $|G_M|^2$.}
	\renewcommand\arraystretch{1.9}
	\begin{tabular}{c|c|c}
		\hline
		\multirow{1}*\text{Mode} & $|G_E|^2(10^{-3})$ & $|G_M|^2(10^{-3})$ \\
		\hline
		$p\bar{p}$ & $0.238\pm0.018$ & $1.87\pm0.04$ \\
		$n\bar{n}$ & $ 0.30\pm0.17$ & $ 1.77\pm0.23 $  \\
		$\Sigma^{+}\bar{\Sigma}^{-}$ & $0.91\pm0.19 $ & $ 0.59\pm0.13$ \\
		$\Sigma^{0}\bar{\Sigma}^{0}$ & $0.661\pm0.025 $ & $0.503\pm0.023$ \\
		$\Lambda\bar{\Lambda}$ & $0.303\pm0.010 $ & $1.640\pm0.032 $  \\
		$\Xi^{0}\bar{\Xi}^{0}$ & $0.108\pm0.028 $ & $1.06\pm0.05$\\
		$\Xi^{-}\bar{\Xi}^{+}$ & $0.12\pm0.04$ & $0.92\pm0.08$ \\
		\hline
	\end{tabular}
\end{table}

\begin{table}
	\caption{Amplitudes parameterization.}
	\label{table:resultAmp} 
	\renewcommand\arraystretch{1.9}
	\begin{tabular}{c|c}
		\hline
		\multirow{1}*\text{Mode} & \text{Amplitudes}  \\
		\hline
		$p \overline p$ & $(G_0 - D_m + F_m)(1+R) e^{i \varphi} + D_e + F_e$ \\
		$n \overline n$ & $(G_0 - D_m + F_m) e^{i \varphi} - 2\,D_e$ \\
		$\Sigma^+ \overline \Sigma^-$ & $(G_0 + 2 D_m)(1+R) e^{i \varphi} + D_e + F_e$ \\
		$\Sigma^0 \overline \Sigma^0$ & $(G_0 + 2 D_m) e^{i \varphi} + D_e$ \\
		$ \Lambda \overline \Lambda$ & $(G_0 - 2 D_m) e^{i \varphi} - D_e$ \\
		$\Xi^0 \overline \Xi^0$ & $(G_0 - D_m - F_m) e^{i \varphi} - 2\,D_e$ \\
		$\Xi^- \overline \Xi^+$ & $(G_0 - D_m - F_m)(1+R) e^{i \varphi} + D_e - F_e$ \\
		\hline
	\end{tabular}
\end{table}

\begin{table*}
	\caption{The values of amplitudes parameters.}
	\label{table:resultpart} 
	\renewcommand\arraystretch{1.8}
	\begin{tabular}{c|c|c|c|c|c|c|c}
		\hline
		\multirow{1}*\text{Mode} & $g_0(10^{-3})$ & $d_e(10^{-3})$ & $d_m(10^{-3})$ & $f_e(10^{-3})$ & $f_m(10^{-3})$ & $r$ & $\cos\phi$ \\
		\hline
		\multirow{1}*\text{the electric part}  & $0.132\pm0.009$ & $0.097\pm0.068$ & $0.368\pm0.021$ & $-0.323\pm0.106$ & $-0.408\pm0.123$ & $-0.081\pm0.104$ & $-0.612\pm0.152$\\ 
		\hline
		\multirow{1}*\text{the magnetic part} & $0.957\pm0.062$ & $0.262\pm0.217$ & $-0.165\pm0.069$ & $0.130\pm0.149$ & $0.168\pm0.041$ & $-0.021\pm0.154$ & $0.124\pm0.359$ \\ 
		\hline
	\end{tabular}
\end{table*}

\section{Summary} \label{sec:summary}

In this article, we use the covariant L-S scheme for the PWA to construct the coupling of $\psi(1^-) \to B_8(1/2^+) \bar{B}_8(1/2^-)$ and fit experiment's measured parameters to calculate the effective radii of these processes. 
For the processes $J/\psi\to\Lambda\bar{\Lambda}$, $J/\psi\to\Sigma^+\bar{\Sigma}^-$ and $\psi(2S)\to\Sigma^+\bar{\Sigma}^-$, the effective radii are deduced to be around 0.04 fm.
However, for other processes of $\psi\to B_8(1/2^+) \bar{B}_8(1/2^-)$, only the range of the effective radii are computed since there is only one measured parameter $\alpha$. 
To fix the radii of these processes, we give some proposals.
Especially, for the $\psi\to p\bar{p}$ or $n\bar{n}$, the polarization observables could help to extract the effective radii of them.
From our calculation, the effective radius of $\psi$ decay is much smaller than the normal size of nucleon, therefore, the angular momenta between baryon and anti-baryon are very small.
Thus, only excited resonances with low spins can be produced and observed in the $\psi$ decay. 
It hence provides a unique window to look for low spin nucleon and hyperon resonances with masses above 2 GeV in the $\psi$ decay.

\section*{Acknowledgments}
We thank useful discussions and valuable comments from Fengkun Guo, Xiaorui Lyu, Ronggang Ping, Jifeng Hu and Jianbin Jiao. 
This work is supported by the NSFC and the Deutsche Forschungsgemeinschaft (DFG, German Research
Foundation) through the funds provided to the Sino-German Collaborative
Research Center TRR110 “Symmetries and the Emergence of Structure in QCD”
(NSFC Grant No. 12070131001, DFG Project-ID 196253076 - TRR 110), by the NSFC 
Grant No.11835015, No.12047503, and by the Chinese Academy of Sciences (CAS) under Grant No.XDB34030000, 
also by the Fundamental Research Funds for the Central Universities (J.J.W), and National Key R$\&$D Program of China under Contract No. 2020YFA0406400 (J.J.W).

\begin{appendix}

\section{THE COVARIANT L-S SCAME}\label{AppendixA}
In this appendix we briefly introduce the covariant L-S scheme for the effective $N^*NM$ couplings.

For a hadron deacy $a\to b + c$, the conservation relation of total angular momentum lead to: 
\begin{align}
\boldsymbol{S}_a = \boldsymbol{S}_{bc}+\boldsymbol{L}_{bc},
\end{align}
where $\boldsymbol{L}_{bc}$ and $\boldsymbol{S}_{bc}$ is the relative orbital angular momentum and the total spin between the particle b and c. 

Then we separate first the orbital angular momentum wave function $\tilde{t}^{{\mu}}$ and covariant spin wave functions $\psi(\Psi)$ or $\phi(\Phi)$.

The orbital angular momentum $L_{bc}$ state can be represented by covariant tensor wave functions $\tilde{t}_{\mu_{1}\cdots\mu_{L}}^{(L)}$\cite{Zou:2002yy,Dulat:2011rn,Anisovich:2004zz,Chung:1993da},
\begin{align}
\tilde{g}_{\mu\nu}(p) \equiv g_{\mu\nu} - \frac{p_\mu p_\nu}{p^2} = -\sum_{S=\pm1,0}\epsilon_\mu(p,S)\epsilon_\nu^\star(p,S),
\end{align}
\begin{align}\label{eq:t0}
\tilde{t}^{(0)}=1,
\end{align}
\begin{align}\label{eq:t1}
\tilde{t}^{(1)}_{\mu}=\tilde{g}_{\mu\nu}(p_{a})r^{\nu}\equiv\tilde{r}_{\mu},
\end{align}
\begin{align}\label{eq:t2}
\tilde{t}^{(2)}_{\mu\nu}=\tilde{r}_{\mu}\tilde{r}_{\nu}&-\frac{1}{3}(\tilde{r}\cdot\tilde{r})\tilde{g}_{\mu\nu},\\
&...\nonumber
\end{align}
where $r=p_b-p_{c}$ is the relative four momentum of the two decay products in the parent particle rest frame.

For the covariant spin wave functions, we use function $\psi(\Psi)$ or $\phi(\Phi)$ to describe with four cases $M\to B^*\bar{B}$, $M\to \bar{B}^*B$, $B^*\to B M$ and $\bar{B}^*\to \bar{B} M$.\cite{Zou:2002yy,Dulat:2011rn,Chung:1993da}

For the case of $M\to B^*(n+\frac{1}{2})\bar{B}(\frac{1}{2})$ and $M\to \bar{B}^*(n+\frac{1}{2})B(\frac{1}{2})$, $S_{bc}$ can be n or (n+1):

\begin{align}
\psi^{(n)}_{\mu_{1}\cdots\mu_{n}}&=\bar{u}_{\mu_{1}\cdots\mu_{n}}(p_{b},S_{b})\gamma_{5}v(p_c,S_c),\\[4pt]	
\Psi^{(n+1)}_{\mu_{1}\cdots\mu_{n+1}}
&=\bar{u}_{\mu_{1}\cdots\mu_{n}}(p_{b},S_{b})\bar{\gamma}_{\mu_{n+1}}v(p_c,S_c)\nonumber\\
&+(\mu_{1}\leftrightarrow\mu_{n+1})+\cdots+(\mu_{n}\leftrightarrow\mu_{n+1}),
\end{align}
\begin{align}
\psi^{C(n)}_{\mu_{1}...\mu_{n}}&=-\bar{u}(p_{c},S_{c})\gamma_{5}v_{\mu_{1}...\mu_{n}}(p_{b},S_{b}),\\[4pt]
\Psi^{C(n+1)}_{\mu_{1}...\mu_{n+1}}
&=\bar{u}(p_{c},S_{c})\bar{\gamma}_{\mu_{n+1}}v_{\mu_{1}...\mu_{n}}(p_{b},S_{b})\nonumber\\
&+(\mu_{1}\leftrightarrow\mu_{n+1})+...+(\mu_{n}\leftrightarrow\mu_{n+1}),
\end{align}
where $\bar{\gamma}_{\mu}\equiv\gamma_{\mu} - r_{\mu}/(m_{a}+m_{b}+m_{c})$.

And use the orbital angular momentum covariant tensors $\tilde{t}_ {\mu_{1} \cdots \mu_{L}}^{(L)}$, covariant spin wave functions $\psi(\Psi)$, metric tensor $g_{\mu\nu}$, totally antisymmetric Levi-Civita tensor $\epsilon_{\mu\nu\lambda\sigma}$ and momentum of the parent particle $p_\mu$ we can get the effective couplings.

\section{CALCULATING THE TWO FORMS OF AMPLITUDE}\label{AppendixB}
In this appendix we will derive the relationship between the two parameters in Eq.~(\ref{eq:tran}).
\begin{align}
A_S&=\Psi^{(1)}_{\mu}\epsilon^{\mu}=\bar{u}_{B}(\gamma_{\mu}-\frac{r_{\mu}}{m_\psi + 2m_B})v_{\bar{B}}\epsilon^{\mu},\\[4pt]
A_D &= \Psi^{(1)}_{\mu}\epsilon_{\nu}\tilde{t}^{(2)\mu\nu}=\bar{u}_{B}(\gamma^{\nu} -  \frac{r^{\nu}}{m_\psi + 2m_B})\nonumber\\
&\cdot(r_{\mu}r_{\nu} - \frac{r^2}{3}g_{\mu\nu} + \frac{r^2}{3m_\psi^2}p^\psi_{\mu}p^\psi_{\nu})v_{\bar{B}}\epsilon^{\mu}\\
=&\bar{u}_{B}[-\frac{r^2}{3}\gamma_{\mu} + (2m_B - \frac{2}{3}\frac{r^2}{m_\psi + 2m_B})r_\mu]v_{\bar{B}}\epsilon^{\mu}.
\end{align}

So,
\begin{align}
A&=g_{S}\bar{u}_{B}[(1 - \frac{r^2}{3}\boldsymbol{y})\gamma_{\mu}- (\frac{1}{m_\psi + 2m_B} \nonumber\\
& - 2m_B\boldsymbol{y} + \frac{2}{3}\frac{r^2}{m_\psi + 2m_B}\boldsymbol{y})r_\mu]v_{\bar{B}}\epsilon^{\mu}\label{eq:amp11},\\[4pt]
\tilde{A}&=-ie_{g}G_M^{\psi}\bar{u}_B[\gamma_{\mu}-\frac{2m_B}{r^2}(1-\boldsymbol{x})r_\mu]v_{\bar{B}}\epsilon^{\mu},\label{eq:amp12}
\end{align}
where $\boldsymbol{y}\equiv e^{i\delta}g_{D}/g_{S}$, $\boldsymbol{x}\equiv G_{E}^{\psi}/G_{M}^{\psi}$.

Compared with Eqs.~(\ref{eq:amp11},\ref{eq:amp12}),we can get:
\begin{align}
&(1 - \frac{r^2}{3}\boldsymbol{y})\frac{2m_B}{r^2}(1-\boldsymbol{x})\nonumber\\
 =& \frac{1}{m_\psi + 2m_B} - 2m_B\boldsymbol{y} + \frac{2}{3}\frac{r^2}{m_\psi + 2m_B}\boldsymbol{y}.
\end{align}

So, the parameters in two forms of amplitude must satisfy:
\begin{align}
\boldsymbol{y}=-\frac{3}{2r^2}\frac{2m_B(m_\psi+2m_B)-r^2-2m_B(m_\psi+2m_B)\boldsymbol{x}}{2m_B(m_\psi+2m_B)-r^2+m_B(m_\psi+2m_B)\boldsymbol{x}}.\nonumber
\end{align}

It equals Eq.~(\ref{eq:tran}).

\end{appendix}

\bibliographystyle{plain}

\end{document}